\documentclass[preprint,aps]{revtex4-1}
\pdfoutput=1

\usepackage{amsmath,amssymb,amsfonts,dcolumn,color,graphicx,graphics,latexsym,placeins,epsfig}
\usepackage{epsfig}
\usepackage{bm}
\usepackage{slashed}
\usepackage{latexsym}
\usepackage{natbib}
\usepackage{url}
\usepackage{dcolumn}
\usepackage{color}
\usepackage{amsfonts,amssymb,amsmath}
\usepackage{graphicx,epsfig}
\usepackage{psfrag}
\usepackage{subfigure}
\usepackage{tabularx}
\usepackage{hyperref}
\hypersetup{colorlinks=true}

\newcommand{\be}{\begin{equation}}
\newcommand{\ee}{\end{equation}}
\newcommand{\ba}{\begin{eqnarray}}
\newcommand{\ea}{\end{eqnarray}}

\begin{document}

\title{Constraints on axionic fuzzy dark matter from light bending and Shapiro time delay }
%\title{Constraints on axion parameters from compact binary systems and bound on the initial misallignment angle for the axionic dark matter in string theory}
\author{Tanmay Kumar Poddar}
\email[Email Address: ]{tanmay@prl.res.in}
%\author{Subhendra Mohanty$^{1}$\footnote{mohanty@prl.res.in}}
%\author{Tanmay Kumar Poddar\footnote{tanmay@prl.res.in}}
%\author{Soumya Jana$^{1,2}$\footnote{Soumya.Jana@etu.unige.ch}}
%\author{Subhendra Mohanty\footnote{mohanty@prl.res.in}}
% \author{A. R. Prasanna$^{b,}$\footnote{prasanna@prl.res.in}}
\affiliation {\it Theoretical Physics Division, Physical Research Laboratory, Ahmedabad 380009, India}
\affiliation{Discipline of Physics, Indian Institute of Technology, Gandhinagar - 382355, India}

%\affiliation{${}^{2}$ {\it D\'epartement de Physique Th\'eorique, Universit\'e de Gen\`eve, 24 quai Ernest Ansermet, 1211Gen\`eve 4, Switzerland} }

\begin{abstract}
Ultralight axion like particles (ALPs) of mass $m_a\in (10^{-21}\rm{eV}-10^{-22}\rm{eV})$ with axion decay constant $f_a\sim 10^{17}\rm{GeV}$ can be candidates for fuzzy dark matter (FDM). If celestial bodies like Earth and Sun are immersed in a low mass axionic FDM potential and if the ALPs have coupling with nucleons then the coherent oscillation of the axionic field results a long range axion hair outside of the celestial bodies. The range of the axion mediated Yukawa type fifth force is determined by the distance between the Earth and the Sun which fixes the upper bound of the mass of axion as $m_a\lesssim10^{-18}\rm{eV}$. The long range axionic Yukawa potential between the Earth and Sun changes the gravitational potential between them and contribute to the light bending and the Shapiro time delay. From the observational uncertainties of those experiments, we put an upper bound on the axion decay constant as $f_a\lesssim 9.85\times 10^{6}\rm{GeV}$, which is the stronger bound obtained from Shapiro time delay. This implies if ALPs are FDM, then they do not couple to nucleons.
\end{abstract}

\maketitle
\section{Introduction}
\label{sec1}
The galactic rotation curve \cite{Rubin:1980zd,Persic:1995ru} and the Bullet cluster experiment by Chandra X-ray observatory \cite{Clowe:2003tk} confirms the existance of non luminous /dark matter (DM) in our universe which constitutes $25\%$ of the total energy budget and cannot be explained by our known standard model (SM) of particle physics \cite{Ade:2015xua}. The model of weakly interacting massive particles (WIMPs) motivated from supersymmetric theory (SUSY) is a promising candidate of DM \cite{Jungman:1995df}. However, direct detection experiments put stringent bound on the WIMPs of mass $\gtrsim 1\rm{GeV}$ \cite{Akerib:2013tjd,Cui:2017nnn,Aprile:2018dbl}. The other problem with WIMP model is that it cannot explain the small scale structure problem of the universe \cite{Moore:1994yx,Oh:2015xoa}. To resolve these drawbacks, physicists think of alternative models like feebly interacting massive particles (FIMPs) \cite{Hall:2009bx}, strongly interacting massive particles (SIMPs) \cite{Hochberg:2014dra}, fuzzy dark matter (FDM) \cite{Hu:2000ke} etc., where particles such as sterile neutrino \cite{Boyarsky:2018tvu}, axions or axion like particles \cite{Duffy:2009ig,Poddar:2019zoe}, ultralight particles \cite{Hui:2016ltb,Poddar:2019wvu,Poddar:2020exe}, primordial black holes \cite{Carr:2016drx,Lacki:2010zf} etc., can be the possible dark matter candidates. These DM particles can have a wide mass range varying from a very few $\rm{eV}$ to several $\rm{GeV}$. 

In this paper, we consider FDM model where the mass of the particle is $\mathcal{O}(10^{-21}\rm{eV}-10^{-22}\rm{eV})$ and it can solve the cuspy halo problem. This ultralight dark matter particle has a de Broglie wavelength of the size of a dwarf galaxy $(1-2\rm{kpc})$ and can form a Bose-Einstein condensate. Ultralight scalar or vector particles, axions or axion like particles (ALPs) can be the candidates of FDM. In the following, we have considered the ultralight ALPs as FDM candidates.

The main motivation of introducing axions in nature was to solve the strong CP problem and it was first proposed by Peccei and Quinn in 1977 \cite{Peccei:1977hh,Weinberg:1977ma,Wilczek:1977pj,Peccei:1977ur}. The direct experimental probe of strong CP problem is the measurement of neutron electroc dipole moment (nEDM). The nEDM depends on a parameter $\bar{\theta}$ which is related to the quantum chromodynamics (QCD) $\theta$ angle by $\bar{\theta}=\theta +\rm{arg}(\rm{det}(M_q))$ where $M_q$ is the quark mass matrix \cite{Adler:1969gk,Bell:1969ts}. From chiral perturbation theory, we can write the nEDM as $d_n\sim 10^{-16}\bar{\theta}e.\rm{cm}$. The experimental bonud on nEDM $(d_n<10^{-26}e.\rm{cm})$ puts upper bound on nEDM parameter as $\bar{\theta}<10^{-10}$ \cite{Baker:2006ts}. The natural choice of $\bar{\theta}\sim\mathcal{O}(1)$ violates the experimental bound which is called the strong CP problem. To solve this problem of having very small value of $\bar\theta$, Peccei and Quinn (PQ) proposed that $\bar{\theta}$ is not just a parameter but a dynamical field and is driven to zero by its own classical potential. This $\bar{\theta}$ is the axion field which is scaled by $f_a$ (some energy scale) to make $\bar\theta=\frac{a}{f_a}$ dimensionless. The axion is a pseudo nambo Goldstone boson which arises due to spontaneous symmetry breaking of global $U(1)$ PQ symmetry at the scale $f_a$ and explicitly breaks at the QCD scale $(\Lambda_{QCD})$ by non perturbative QCD effects. These are QCD axions and can couple with the other SM particles with interaction strength $\propto\frac{1}{f_a}$ \cite{Profumo:2019ujg}. Hence, larger values of $f_a$ implies weaker coupling with matter. There are other ultralight pseudoscalar particles which are not exactly the QCD axions but have similar kind of interactions. Those particles are called axion like particles or ALPs and are well motivated from string theory \cite{Svrcek:2006yi}. So far, there are no experimental confirmation of the presence of axions however astrophysical, cosmological, laboratory and other experiments put bounds on the axion parameters \cite{Inoue:2008zp,Arik:2008mq,Hannestad:2005df,Melchiorri:2007cd,Hannestad:2008js,Hamann:2009yf,Semertzidis:1990qc,Cameron:1993mr,Robilliard:2007bq,Chou:2007zzc,Sikivie:2007qm,Kim:1986ax,Cheng:1987gp,Rosenberg:2000wb,Hertzberg:2008wr,Visinelli:2009zm,Battye:1994au,Yamaguchi:1998gx,Hagmann:2000ja}. The axions can also be probed from superradiance \cite{Plascencia:2017kca,Chen:2019fsq} and birefringence phenomena \cite{Liu:2019brz,Sigl:2018fba,Poddar:2020qft}. The ultralight ALPs with mass $m_a\sim 10^{-22}\rm{eV}$ and $f_a\sim10^{17}\rm{GeV}$ satisfy the cold FDM relic density which are produced from vacuum misalignment mechanism \cite{Hui:2016ltb}.

In a macroscopic unpolarized body, if ALPs have spin dependent coupling with nucleons then there is no net long range force due to ALPs outside the body. However, if they have CP violating coupling then they can mediate long range force even for unpolarized body \cite{Moody:1984ba,Raffelt:2012sp}.

It has been proposed in \cite{Hook:2017psm} that if compact objects like neutron star (NS), white dwarf (WD), and celestial bodies like Sun, Earth etc., are immersed in a low mass axionic FDM potential and if the axions have coupling with nucleons then the coherent oscillation of the axionic field results a long range axion hair outside of those objects. The long range Yukawa type of axionic potential between Sun and Earth changes the effective gravitational potential and affects in the measurement of bending of light and Shapiro time delay.

The bending of light or the gravitational lensing \cite{Will:2014kxa,Will:2014zpa} is one of the tests of Einstein's general theory of relativity (GR) along with the perihelion precession of Mercury planet and the gravitational redshift \cite{Einstein:1916vd}. When light ray from a distant star passes through a massive object like Sun then the speed of light decreases due to the presence of increasing gravitational potential. In other words, massive objects with higher gravity distorts the spacetime geometry and bends the light. In 1915, Einstein became the first person to calculate the amount of bending of light near the Sun which is $1.75 \rm{arcsec}$ based on equivalance principle. This value agrees well with the experiment to an uncertainty of $\sim 10^{-4}$ \cite{Fomalont:2009zg}. Another test of Einstein's GR theory is the Shapiro time delay which was predicted by Irwin Shapiro in 1964 \cite{Shapiro:1964uw,Shapiro:1968zza}. When a radar signal is sent from Earth to Venus and it reflects back from Venus to Earth, then the time taken for the round trip is delayed by the presence of strong gravitational potential near the Sun. The calculated amount of time delay is $2\times 10^{-4}\rm{sec}$ which agrees well with the experiment to an uncertainty of $\sim 10^{-5}$ \cite{Bertotti:2003rm}. Gravitational waves, high energy neutrinos etc., also have this Shapiro time delay from which one can constrain the violation of weak equivalence principle \cite{Kahya:2016prx,Boran:2018ypz}.

The Earth and Sun which are the sources of axions can mediate a long range Yukawa type of potential and result an axionic fifth force between those massive objects. This long range Yukawa potential affects the effective gravitational potential between Earth and Sun and contribute to the bending of light and Shapiro time delay within the experimental uncertainty.
 
It has been studied in \cite{Poddar:2019zoe}, that if axions are sourced by NS and WD, then long range axion hair can mediate between NS-NS and NS-WD binary systems and contribute to the orbital period loss. From the observational uncertainty of the orbital period decay the authors of \cite{Poddar:2019zoe} put bound on the axion parameters. In this paper we calculate the light bending and Shapiro time delay due to the presence of long range axionic fifth force between Earth and Sun and put bounds on the axion mass and axion decay constant. 

The paper is organized as follows. In Sec.\ref{sec2}, we have discussed the long range behaviour of the axion field and the axion charge for a massive object immersed in an ultralight axion potential. In Sec.\ref{sec2a}, we have explained how the Earth and the Sun can be the possible sources of axions. In Sec.\ref{sec3} and Sec.\ref{sec4} we have calculated the amount of light bending and Shapiro time delay due to the long range axionic fifth force. We put bounds on the axion parameters in Sec.\ref{sec5} from the observational uncertainty of light bending and Shapiro delay. In Sec.\ref{sec6}, we put constraints on axionic FDM. Finally, in Sec.\ref{sec7}, we conclude our result.

In the rest of the paper, we use natural system of units $\hbar=c=1$ and $G=1$.

The parameters that we have chosen in our following analysis are: the radius of the Sun $R_\odot\sim r_0\sim b=6.96\times 10^{10}\rm{cm}=3.52\times 10^{24}\rm{GeV^{-1}}$, the radius of the Earth $R_\oplus=6.38\times 10^8\rm{cm}=3.22\times 10^{22}\rm{GeV^{-1}}$, the distance between Earth and Sun is $D=r_e=1.49\times 10^{13}\rm{cm}=7.52\times 10^{26}\rm{GeV^{-1}}$, the distance between Sun and Venus is $r_v=1.08\times 10^{13}\rm{cm}=5.47\times 10^{26}\rm{GeV^{-1}}$, the mass of Sun $M=M_\odot=10^{57}\rm{GeV}$, the mass of Earth $M_p=M_\oplus=3.35\times 10^{51}\rm{GeV}$, $G=10^{-38}\rm{GeV^{-2}}$.
\section{The axion profile for a compact/celestial object}
\label{sec2}
The Lagrangian which describes the interaction of ALPs with other SM particles below the PQ and the electroweak breaking scale in the leading order of $\frac{1}{f_a}$ is
\begin{equation}
\mathcal{L}=\frac{1}{2}\partial_\mu a \partial^\mu a-\frac{\alpha_s}{8\pi}g_{ag}\frac{a}{f_a}G^{\mu\nu}_a\tilde{G}^a_{\mu\nu}-\frac{\alpha}{8\pi}g_{a\gamma}\frac{a}{f_a}F^{\mu\nu}\tilde{F}_{\mu\nu}+\frac{1}{2f_a}g_{af}\partial_\mu a\bar{f}\gamma^\mu\gamma_5f,
\label{eq:b1}
\end{equation}
where $g's$ denote the coupling constants which depend on the model. The first term in Eq.(\ref{eq:b1}) denotes the kinetic term of the dynamical axion field, whereas the second, third and fourth terms denote the interactions of axion with gluon, photon and fermion fields respectively. All the coupling terms in Eq.(\ref{eq:b1}) are proportional to $\frac{1}{f_a}$ which means larger value of $f_a$ leads to weaker coupling of matter with axions. 

It has been discussed in \cite{Hook:2017psm} that if ALPs are coupled with nucleons then massive objects like Sun, Earth, neutron stars, white dwarfs etc., can be the sources of long range axion hair. Here we consider the massive objects as Sun and Earth. In vacuum, the ALPs potential is given by
\begin{equation}
V\approx-\epsilon m^2_\pi f^2_\pi\sqrt{1-\frac{4m_um_d}{(m_u+m_d)^2}\sin^2\Big(\frac{a}{2f_a}\Big)},
\label{eq:b2}
\end{equation}
where $m_\pi$ and $f_\pi$ are the pion mass and pion decay constant respectively. $m_u$ and $m_d$ are the masses of up and down quarks, $a$ denotes the axion field and $\epsilon$ is a small number which is fixed by the axion mass that we want to probe. We chose $m_u=m_d$ for convenience and the ALPs mass in vacuum becomes
\begin{equation}
m_a=\frac{m_\pi f_\pi}{2f_a}\sqrt{\epsilon}\hspace{0.5cm}r>R,
\label{eq:b3}
\end{equation}
where $R$ is the radius of the massive object. Now, inside the massive object, the ALPs potential is
\begin{equation}
V\approx -m^2_\pi f^2_\pi\Big\{\Big(\epsilon-\frac{\sigma_Nn_N}{m^2_\pi f^2_\pi}\Big)\Big|\cos\Big(\frac{a}{2f_a}\Big)\Big|+\mathcal{O}\Big(\Big(\frac{\sigma_Nn_N}{m_\pi^2f_\pi^2}\Big)^2\Big)\Big\},
\label{eq:b4}
\end{equation}
where the nucleon density $n_N$ corrects the quark mass $m_q$ which is denoted by $\frac{\sigma_Nn_N}{m^2_\pi f^2_\pi}$ and changes the ALPs potential. The nucleon $\sigma$ term $\sigma_N$ is defined by
\begin{equation}
\sigma_N=\sum_{q=u,d}m_q\frac{\partial m_N}{\partial m_q}.
\label{eq:b5}
\end{equation}
Inside the massive object, the ALPs mass is tachyonic and its magnitude is given by
\begin{equation}
m_T=\frac{m_\pi f_\pi}{2f_a}\sqrt{\frac{\sigma_Nn_N}{m_\pi^2f_\pi^2}-\epsilon}, \hspace{0.5cm} r<R.
\label{eq:b6}
\end{equation}
Inside the massive object, $\sigma_N\neq 0$ and $m_T\gtrsim m_a$. The high nucleon density inside the massive object changes the sign of ALPs potential which allows the objects as the sources of ALPs. The axion potential is periodic and it has a degenerate vacuua which can be weakly broken by finite density effect or higher dimensional operators suppressed at the Planck scale. 

Inside the massive object $(r<R)$ $\sigma_N\neq0$, and the ALPs potential attains maxima at $a=0, \pm 4\pi f_a\cdots$ and minima at $a=\pm 2\pi f_a, \pm 6\pi f_a\cdots$. Outside the massive object $(r>R)$, $\sigma_N=0$ and the ALPs potential have maxima at the field values $a=\pm 2\pi f_a, \pm 6\pi f_a\cdots$ and minima at the field values $a=0, \pm 4\pi f_a\cdots$.

In the region $r<R$, the axion field sits on one of the local maxima of the axion potential. In the region $r>R$, the field rolls down to the nearest local minimum of the potential and stabilizes about it. Inside the massive object, the axion field takes a constant value $a=4\pi f_a$, the nearest local maximum and reaches $a=0$ asymptotically at $r\rightarrow\infty$ and allows ALPs to be sourced by the object. This happens due to the fact that the gain in potential energy $m^2_\pi f^2_\pi(\epsilon-\frac{\sigma_Nn_N}{m_\pi^2f_\pi^2})$ which is obtained by putting $a=4\pi f_a$ in Eq.(\ref{eq:b4}) is greater than the gradient energy $\frac{f^2_a}{r^2}$ which is required to move the axion from its unstable solution. This implies
\begin{equation}
r_{critical}\gtrsim \frac{1}{m_T},
\label{eq:b7}
\end{equation}
where $r_{critical}$ is the critical size and if the size of the compact object is greater than $r_{critical}$, then ALPs can be emiited from those objects. We can obtain the long range behaviour of the axion field by matching the inside and outside axion field solution. Now the equation of motion of axion field for a massive object of constant density is \cite{Hook:2017psm}
\begin{equation}
\nabla^{\mu}\nabla_{\mu} \left(\frac{a}{2f_a}\right)= \begin{cases} 
     -m_T^2\sin \left(\frac{a}{2f_a}\right) \text{sgn}\lbrace \cos \left(\frac{a}{2f_a}\right)\rbrace & (r<R),\\
       m_a^2\sin \left(\frac{a}{2f_a}\right) \text{sgn}\lbrace \cos \left(\frac{a}{2f_a}\right)\rbrace & (r>R).
   \end{cases}
   \label{eq:b8}
   \end{equation}
The solution of Eq.(\ref{eq:b8}) in the Schwarzschild background is \cite{Poddar:2020qft}
\begin{eqnarray}
a&=&\frac{q_a e^{-m_ar}}{r}\Big[1+\frac{M}{r}\{1-m_ar\ln(m_ar)+m_are^{2m_ar}Ei(-2m_ar)\}\Big]+\mathcal{O}\Big(\Big(\frac{M}{R}\Big)^2\Big),
r>R.\nonumber\\
&=&4\pi f_a, \hspace{11.0cm} \hspace{1.0cm}r<R,
\label{eq:b9}
\end{eqnarray}
where we solve Eq.(\ref{eq:b8}) in a perturbative way, $\frac{M}{R}$ is the perturbation parameter, the leading order term is the Yukawa term and $q_a$ denotes the axion charge which is given as \cite{Poddar:2020qft}
\begin{equation}
q_a=4\pi f_a R e^{m_aR}\Big[1+\frac{M}{R}\{1-m_a R \ln(m_aR)+m_aRe^{2m_aR}Ei(-2m_aR)\}\Big]^{-1}+\mathcal{O}\Big(\Big(\frac{M}{R}\Big)^{-2}\Big).
\label{eq:10}
\end{equation}
In the limit $\frac{M}{R}\ll1$ and $m_a\rightarrow 0$, we obtain $q_a\sim 4\pi f_a R$ and $a(r>R)\sim\frac{q_ae^{-m_ar}}{r}$.

\begin{figure}[!htbp]
\centering
\subfigure[$V$ vs. $a$]{\includegraphics[width=3.0in,angle=360]{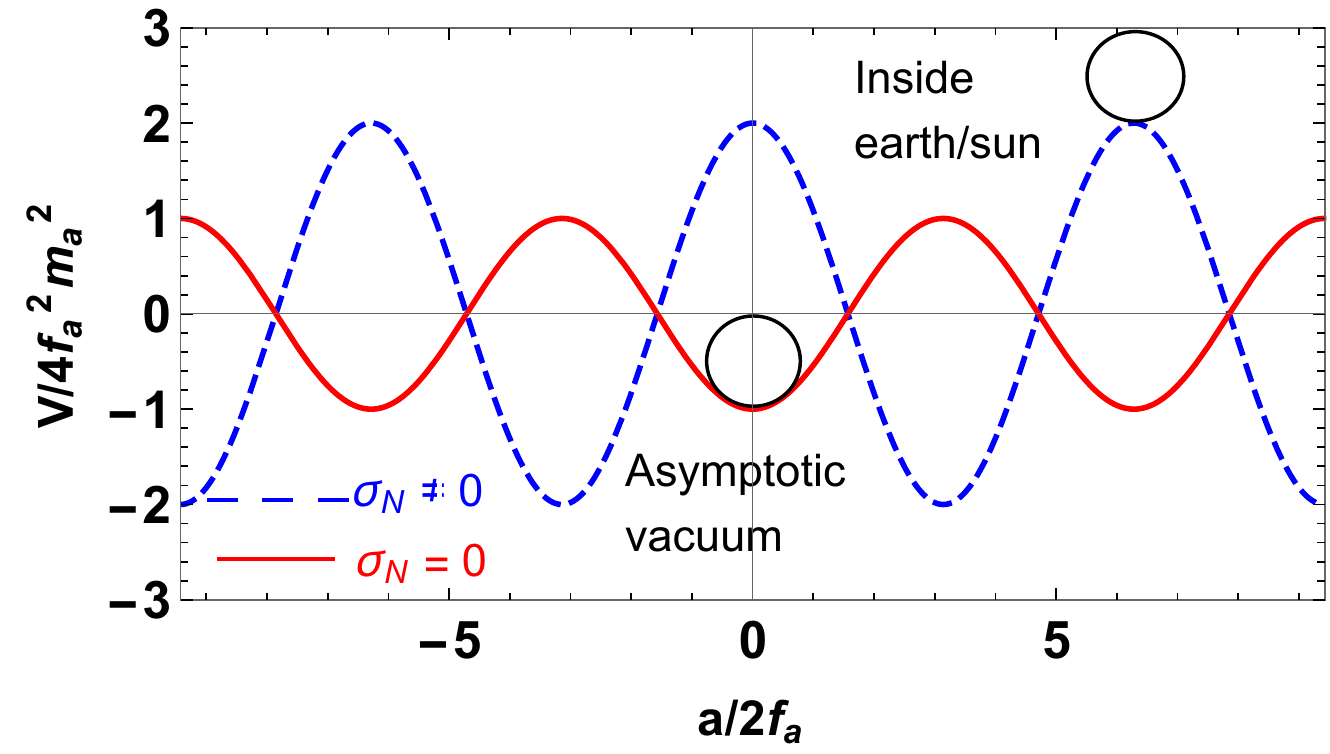}\label{subfig:vvsa}}
\subfigure[$a$ vs. $r$]{\includegraphics[width=3.0in,angle=360]{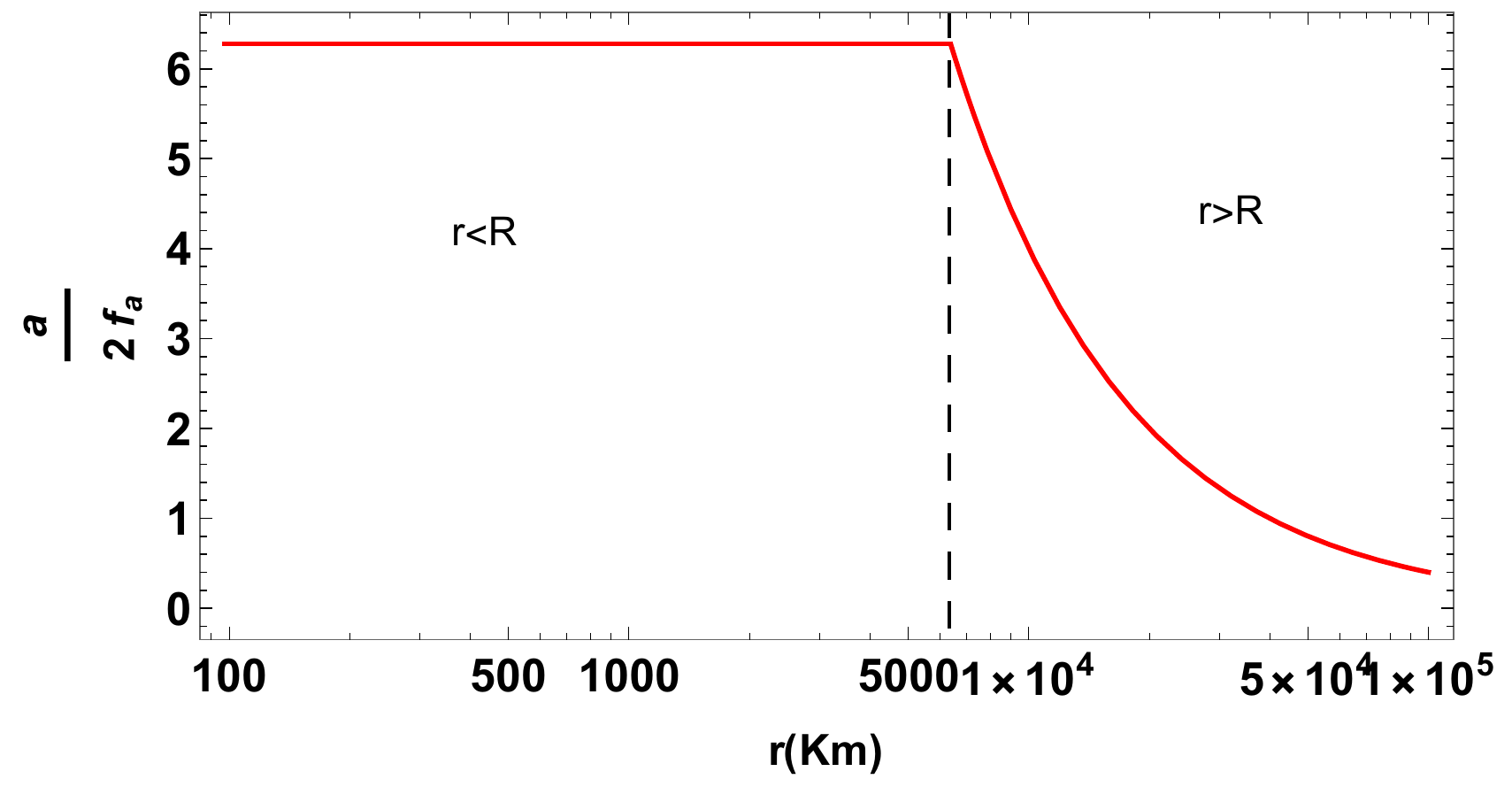}\label{subfig:avsr}}
\subfigure[$V$ vs. $r$]{\includegraphics[width=3.0in,angle=360]{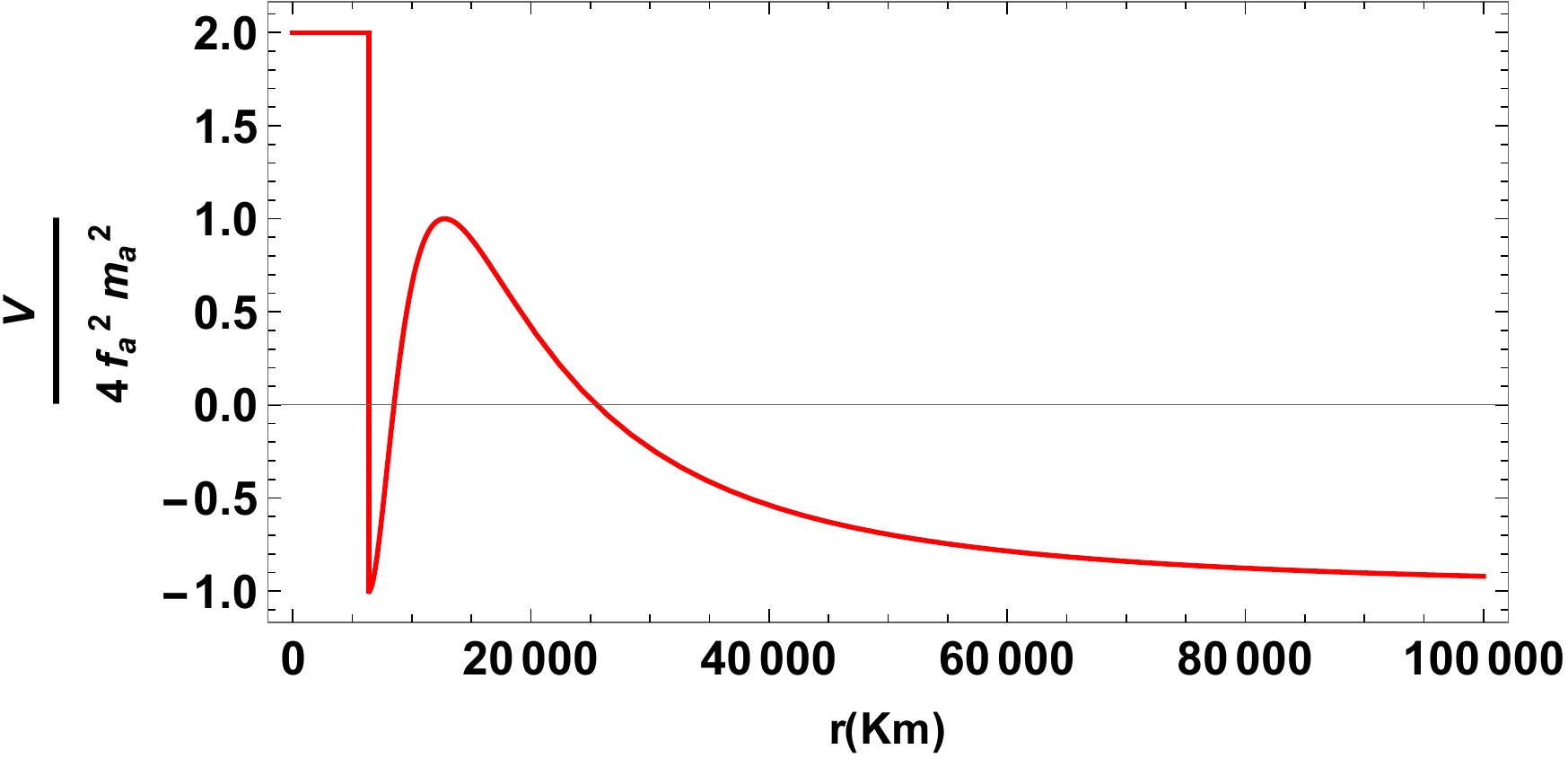}\label{subfig:vvsr}}
\subfigure[$q_a$ vs. $m_a$]{\includegraphics[width=3.0in,angle=360]{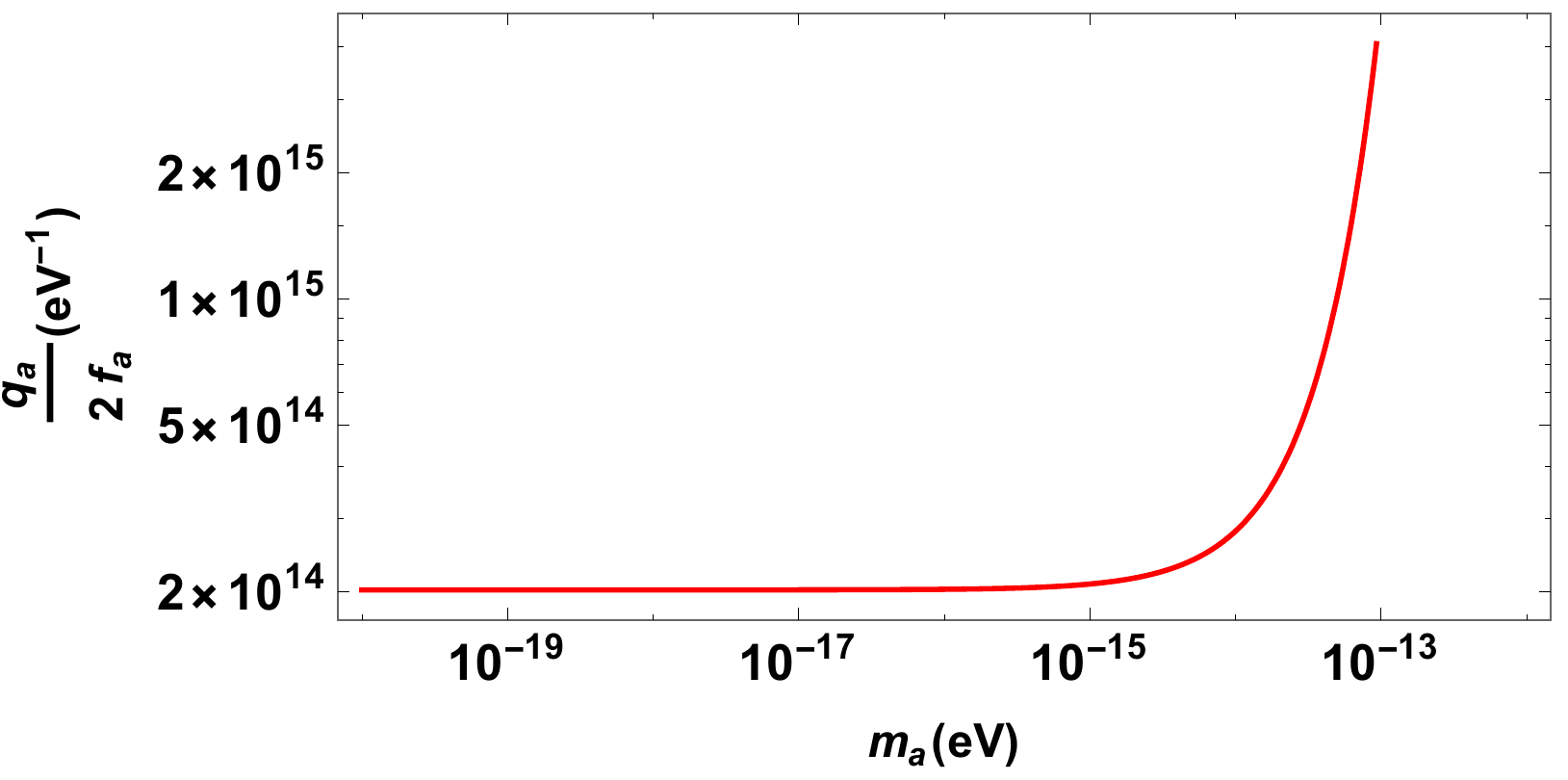}\label{subfig:qavsma}}
\caption{In Fig.\ref{subfig:vvsa}, we plot the variation of the axion potential with the axion field. Here, we have chosen $\frac{m^2_T}{m^2_a}=2$. In Fig.\ref{subfig:avsr}, we have shown the axion field behaviour with distance for Earth. In the figure, $R$ denotes the radius of the Earth. In Fig.\ref{subfig:vvsr} we have shown the variation of the axion potential with the distance. There is a discontinuity of potential at the surface of the Earth due to the sign change of $V$. The variation of axion charge with the mass of axion is shown in Fig.\ref{subfig:qavsma}. We plot the figures in $\frac{M}{R}\ll1$ limit.}
\label{fig:u}
\end{figure}

In Fig.\ref{fig:u}, we have shown the behaviour of axion for Earth. In Fig.\ref{subfig:vvsa}, we have shown the variation of axion potential with the axion field for Earth. Here we have chosen $\frac{m^2_T}{m^2_a}=2$. Inside the Earth, the axion field takes a constant value $a=4\pi f_a$, the nearest local maximum and reaches zero asymptotically at $r\rightarrow \infty$. In Fig.\ref{subfig:avsr}, we have shown the axion field behaviour with distance. The axion field takes a constant value inside the Earth and has a long range Yukawa type behaviour outside of the Earth. In Fig.\ref{subfig:vvsr} we have shown the variation of the axion potential with the distance. The variation of axion charge with the mass of axion is shown in Fig.\ref{subfig:qavsma}. The similar behaviour of the axion field is also true for Sun. In Fig.\ref{fig:u}, we plot all the figures in $\frac{M}{R}\ll1$ limit since for Earth and Sun, the values of $\frac{M}{R}$ are very small. The similar behaviour of axion for compact objects (NS,WD) are also obtained in \cite{Poddar:2019zoe}. For small ALPs mass, the Compton wavelength of ALPs is greater than the size of the massive object and the ALPs field has a long range behaviour of Yukawa type. In Sec.\ref{sec2a} we quantitatively describe how the Earth and the Sun can be the sources of axions.
\section{Sun and Earth as the sources of ALPs}
\label{sec2a}
A celestial object like Earth or Sun can be the source of axions if its size is greater than the critical size which is given by Eq.(\ref{eq:b7}). From Eq.(\ref{eq:b3}) and Eq.(\ref{eq:b6}), we can write
\begin{equation}
m_T^2=\frac{\sigma_Nn_N}{4f^2_a}-m^2_a.
\label{eq:ba1}
\end{equation}
Using the values of $\sigma_N=59\rm{MeV}$ from lattice simulation \cite{Alarcon:2011zs}, $m_a\lesssim1.333\times10^{-18}\rm{eV}$ and other parameters we obtain the upper bounds on $f_a$ for which the axions can be sourced by Earth and Sun as $f_a\lesssim 1.91\times 10^{13}\rm{GeV}$ and $f_a\lesssim 10^{15}\rm{GeV}$ respectively. The mass of the axion is constrained by the distance between Earth and Sun.

In other words, Earth and Sun can be the sources of axions if the following two conditions are satisfied,
\begin{equation}
\rho_R\gtrsim m^2_af^2_a, \hspace{0.5cm} \frac{1}{R}\lesssim\frac{\sqrt{\rho_R}}{f_a},
\label{eq:ba2}
\end{equation}
where $\rho_R$ is the mass density of the celestial body of radius $R$. We have checked that the $m_a$ and $f_a$ values that we obtain later in Sec.\ref{sec5} satisfy Eq.(\ref{eq:ba2}). Hence, the Sun and the Earth are in fact the sources of axions. If $q_1$ and $q_2$ are the axion charges of Sun and Earth respectively, then the potential energy act between Sun and Earth is $V=\frac{q_1q_2}{4\pi r}e^{-m_a r}$ which is long range Yukawa type. Hence, there is a long range axion mediated fifth force act between the Earth and the Sun. The two massive objects attract each other if $q_1q_2>0$ and repel each other if $q_1q_2<0$.

For Earth $\frac{M}{R}\equiv \frac{GM}{R}=1.04\times 10^{-9}$ and for Sun $\frac{M}{R}\equiv \frac{GM}{R}=2.84\times 10^{-6}$ which are much smaller than unity. Hence, we use the axion charge for Earth and Sun as $q_a=4\pi f_aR$ and the axion field outside of the compact object as $a(r>R)=\frac{q_a e^{-m_ar}}{r}$.

\section{Light bending due to long range axionic Yukawa type of potential in the Schwarzschild spacetime background}
\label{sec3}
The trajectory of light or photon follows null geodesic which is given by
\begin{equation}
g_{\mu\nu}V^\mu V^\nu=0,
\label{eq:a1}
\end{equation}
where $V^\mu=\frac{dX^\mu}{d \lambda}$ is the tangent vector of a curve which is a parametrized path through spacetime $x^\mu(\lambda)$, where $\lambda$ is the affine parameter that varies smoothly and monotonically along the path and $x^\mu=(t,r,\theta,\phi)$ are the coordinates of the Schwarzschild spacetime which is defined by the metric $g_{\mu\nu}$ whose line element is  
\begin{equation}
ds^2=\Big(1-\frac{2M}{r}\Big)dt^2-\Big(1-\frac{2M}{r}\Big)^{-1}dr^2-r^2d\Omega^2,
\label{eq:a2}
\end{equation}
where we put Newton's universal gravitation constant $G=1$ for convenience and $d\Omega^2=d\theta^2+\sin^2\theta d\phi^2$. $M$ is the mass of the Sun outside of which Einstein's field solution is defined. For planar motion $\theta=\frac{\pi}{2}$ and the conserved quantities are $E=\Big(1-\frac{2M}{r}\Big)\dot{t}$ and $L=r^2\dot{\phi}$. $E$ and $L$ are interpreted as the energy per unit mass and the angular momentum per unit mass of the system which are constants of motion. 

Using Eq.(\ref{eq:a1}) and Eq.(\ref{eq:a2}) we can write for null geodesic
\begin{equation}
\Big(1-\frac{2M}{r}\Big)\dot{t}^2-\Big(1-\frac{2M}{r}\Big)^{-1}\dot{r}^2-r^2\dot{\phi}^2=0.
\label{eq:a3}
\end{equation}
Expressions of $L$ and $E$ reduce Eq.(\ref{eq:a3}) to
\begin{eqnarray}
%\begin{align}
\frac{E^2}{2}&=&\frac{\dot{r}^2}{2}+\frac{L^2}{2r^2}\Big(1-\frac{2M}{r}\Big)\nonumber\\
&=& \frac{L^2}{2}\Big(\frac{du}{d\phi}\Big)^2+\frac{L^2u^2}{2}(1-2Mu),
%\end{align}
\label{eq:a4}
\end{eqnarray}
where we use $\dot{r}=\frac{dr}{d\lambda}=\frac{L}{r^2}\frac{dr}{d\phi}$ and $u=\frac{1}{r}$ is the reciprocal coordinate. The right hand side of Eq.(\ref{eq:a4}) is the effective potential of the system. As we have already discussed that the Sun and the Earth can be the sources of axions, the long range axion field mediates a Yukawa type fifth force (in addition to the gravitational force) between the Sun and the Earth which changes the effective potential per unit mass of the system as
\begin{equation}
V_{eff}=\frac{L^2}{2}\Big(\frac{du}{d\phi}\Big)^2+\frac{L^2u^2}{2}(1-2Mu)-\frac{q_1 q_2u}{4\pi M_p }e^{-\frac{m_a}{u}},
\label{eq:a5}
\end{equation}
where $q_1$ and $q_2$ are axion charges of the Sun and the Earth respectively, $m_a$ is the mass of the axion and $M_p$ is the mass of the Earth. Hence, Eq.(\ref{eq:a4}) becomes
\begin{equation}
\frac{E^2}{2}=\frac{L^2}{2}\Big(\frac{du}{d\phi}\Big)^2+\frac{L^2u^2}{2}(1-2Mu)-\frac{q_1 q_2u}{4\pi M_p }e^{-\frac{m_a}{u}}.
\label{eq:a6}
\end{equation} 
Differentiating Eq.(\ref{eq:a6}) with respect to $\phi$, we obtain
\begin{equation}
0=\frac{d^2u}{d\phi^2}+u-3Mu^2-\frac{q_1q_2}{4\pi M_pL^2}e^{-\frac{m_a}{u}}-\frac{q_1q_2m_a}{4\pi M_pL^2u}e^{-\frac{m_a}{u}}.
\label{eq:a7}
\end{equation}
Expanding Eq.(\ref{eq:a7}) upto the leading order of $m_a$, we obtain
\begin{equation}
\frac{d^2u}{d\phi^2}+u=3Mu^2+\frac{q_1q_2}{4\pi M_pL^2}-\frac{q_1q_2m_a^2}{8\pi M_pL^2u^2},
\label{eq:a8}
\end{equation}
where the first term in r.h.s of Eq.(\ref{eq:a8}) arises in Einstein's standard GR calculation which causes the light bending and the last two terms contribute to the uncertainty in light bending measurement from experiment compared with the standard GR result. This arises due to long range axion mediated fifth force between the celestial objects which change the effective potential.

Suppose the solution of the Eq.(\ref{eq:a8}) is $u(\phi)=u_0(\phi)+\Delta u(\phi)$, where $u_0(\phi)$ is the solution for the complementary function of Eq.(\ref{eq:a8}) and $\Delta u(\phi)$ is the solution due to GR correction and the Yukawa contribution. Thus we can write
\begin{equation}
\frac{d^2u_0}{d\phi^2}+u_0=0. 
\label{eq:a9}
\end{equation}
The solution of Eq.(\ref{eq:a9}) is $u_0=\frac{\sin\phi}{b}$, where $b$ is the impact parameter and 
\begin{equation}
\frac{d^2\Delta u}{d\phi^2}+\Delta u=3M\frac{sin^2\phi}{b^2}+\frac{q_1q_2}{4\pi M_pL^2}-\frac{q_1q_2m_a^2b^2}{8\pi M_pL^2\sin^2\phi}.
\label{eq:a10}
\end{equation}
The solution of Eq.(\ref{eq:a10}) is
\begin{equation}
\Delta u(\phi)=\frac{3M}{2b^2}\Big(1+\frac{1}{3}\cos2\phi\Big)+\frac{q_1q_2}{4\pi M_pL^2}-\frac{q_1q_2m_a^2b^2}{8\pi M_pL^2}[\cos\phi \ln|\textrm{cosec}\phi+\cot\phi|-1].
\label{eq:a11}
\end{equation}
Hence, the total solution of Eq.(\ref{eq:a8}) is
\begin{equation}
u=\frac{\sin\phi}{b}+\frac{3M}{2b^2}\Big(1+\frac{1}{3}\cos2\phi\Big)+\frac{q_1q_2}{4\pi M_pL^2}-\frac{q_1q_2m_a^2b^2}{8\pi M_pL^2}[\cos\phi \ln|\textrm{cosec}\phi+\cot\phi|-1].
\label{eq:a12}
\end{equation}
Far from the Sun, $u\rightarrow 0$ as $\phi\rightarrow 0$. Hence, from Eq.(\ref{eq:a12}) we can write the change in the angular coordinate $\phi$ is
\begin{equation}
\delta\phi=\frac{\frac{-2M}{b^2}-\frac{q_1q_2}{4\pi M_pL^2}(1-0.347m_a^2b^2)}{\frac{1}{b}+\frac{q_1q_2m_a^2b^2}{8\pi M_pL^2}}.
\label{eq:a13}
\end{equation}
The contribution to $\delta\phi$ before and after the turning point are equal from symmetry. Hence the total light bending is
\begin{equation}
\Delta\phi=-2\delta\phi=\frac{\frac{4M}{b^2}+\frac{q_1q_2}{2\pi M_pL^2}(1-0.347m_a^2b^2)}{\frac{1}{b}+\frac{q_1q_2m_a^2b^2}{8\pi M_pL^2}}.
\label{eq:a14}
\end{equation}
In absence of long range axion mediated Yukawa type of force $(q_1=q_2=0)$, the deflection of light can be written from Eq.(\ref{eq:a14}) as 
\begin{equation}
\Delta\phi=\frac{4M}{b}=\frac{4GM}{R_\odot c^2}=1.75\hspace{0.1cm} \rm{arcsec},
\label{eq:a16}
\end{equation}
which is the standard GR result. We assume $b\sim R_\odot$ as the solar radius, $c$ is the speed of light in vacuum. We replace $M\rightarrow GM$ and $b\rightarrow R_\odot c^2$ in the last step to write the deflection in SI system of units.
\section{Shapiro time delay due to long range axionic Yukawa type of potential in the Schwarzschild spacetime background}
\label{sec4}
To calculate the Shapiro time delay due to long range Yukawa axion potential, we can write Eq.(\ref{eq:a6}) as
\begin{equation}
\frac{E^2}{2}=\frac{\dot{r}^2}{2}+\frac{L^2}{2r^2}\Big(1-\frac{2M}{r}\Big)-\frac{q_1q_2}{4\pi M_pr}e^{-m_ar},
\label{eq:a17}
\end{equation}
where $\dot{r}=\frac{dr}{d\lambda}=\frac{dr}{dt}\frac{dt}{d\lambda}=\frac{E}{\Big(1-\frac{2M}{r}\Big)}\frac{dr}{dt}$. Thus, Eq.(\ref{eq:a17}) becomes
\begin{equation}
\frac{E^2}{2}=\frac{E^2}{2\Big(1-\frac{2M}{r}\Big)^2}\Big(\frac{dr}{dt}\Big)^2+\frac{L^2}{2r^2}\Big(1-\frac{2M}{r}\Big)-\frac{q_1q_2}{4\pi M_pr}e^{-m_ar}.
\label{eq:a18}
\end{equation}
For the closest approach of light, $\frac{dr}{dt}=0$ at $r=r_0$. Hence, from Eq.(\ref{eq:a18}) we can write
\begin{equation}
\frac{L^2}{E^2}=\Big(1+\frac{q_1q_2e^{-m_ar_0}}{2\pi M_pE^2r_0}\Big)\frac{r^2_0}{\Big(1-\frac{2M}{r_0}\Big)}.
\label{eq:a19}
\end{equation}
In absence of axionic Yukawa potential, Eq.(\ref{eq:a19}) becomes $\frac{L^2}{E^2}=\frac{r^2_0}{\Big(1-\frac{2M}{r_0}\Big)}$ which is the standard result in GR. Hence using Eq.(\ref{eq:a19}), we can write Eq.(\ref{eq:a18}) as
\begin{equation}
\frac{E^2}{2}=\frac{E^2}{2\Big(1-\frac{2M}{r}\Big)^2}\Big(\frac{dr}{dt}\Big)^2+\frac{1}{2r^2}\Big(1-\frac{2M}{r}\Big)\frac{E^2r_0^2}{\Big(1-\frac{2M}{r_0}\Big)}\Big(1+\frac{q_1q_2e^{-m_ar_0}}{2\pi M_p E^2r_0}\Big)-\frac{q_1q_2}{4\pi M_pr}e^{-m_a r}.
\label{eq:a20}
\end{equation}
We can obtain the rate of change of $r$ from Eq.(\ref{eq:a20}) as
\begin{equation}
\frac{dr}{dt}=\Big(1-\frac{2M}{r}\Big)\Big[1-\frac{1}{r^2}\Big(1-\frac{2M}{r}\Big)\frac{r_0^2}{\Big(1-\frac{2M}{r_0}\Big)}\Big(1+\frac{q_1q_2e^{-m_ar_0}}{2\pi M_pE^2r_0}\Big)-\frac{q_1q_2}{2\pi M_pE^2r}e^{-m_ar}\Big]^\frac{1}{2}
\label{eq:a21}
\end{equation}
Hence, using Eq.(\ref{eq:a21}), the time taken by the light to reach from $r_0$ to $r$ is
\begin{eqnarray}
t&=&\int^r_{r_0}\frac{dt}{dr}dr\nonumber\\
&=& \int^r_{r_0}dr\frac{1}{\Big(1-\frac{2M}{r}\Big)}\Big[1-\frac{r^2_0}{r^2}\frac{\Big(1-\frac{2M}{r}\Big)}{\Big(1-\frac{2M}{r_0}\Big)}\Big(1+\frac{q_1q_2e^{-m_ar_0}}{2\pi M_pE^2r_0}\Big)-\frac{q_1q_2}{2\pi M_pE^2r}e^{-m_ar}\Big]^{-\frac{1}{2}}.
\label{eq:a22}
\end{eqnarray}
If there is no mass distribution between Earth and Venus, then we can put $M=0$ in Eq.(\ref{eq:a22}) and the required time becomes
\begin{equation}
\begin{split}
t=t_1=\sqrt{r^2-r^2_0}-\frac{1}{2}\frac{a_0}{r}(-r^2_0+2r^2)-\frac{b_0e^{-c_0r}r^2_0}{48r^4}[-36r^2(-1+c_0r)+r^2_0(6-2c_0r+c^2_0r^2)]+\\
\frac{b_0}{48}(48+36c^2_0r^2_0)Ei(-c_0r)+\mathcal{O}(c_0^3),
\end{split}
\label{eq:a23}
\end{equation}
where $a_0=\frac{q_1q_2e^{-m_ar_0}}{4\pi M_pE^2r_0}$, $b_0=\frac{q_1q_2}{4\pi M_pE^2}$, and $c_0=m_a$. $Ei(x)$ is the exponential integral function which is defined as $Ei(x)=-\int^\infty_{-x}\frac{e^{-t}}{t}dt$.

Now if there is a mass distribution between Earth and Venus then $M\neq 0$ and from Eq.(\ref{eq:a22}) we obtain the required time after expanding and linearising in $M$ as
\begin{equation}
\begin{split}
t=t_2=\sqrt{r^2-r^2_0}+2M \ln\frac{\sqrt{r^2-r^2_0}+r}{r_0}+M\Big(\frac{r-r_0}{r+r_0}\Big)^\frac{1}{2}-\frac{(2M+r_0)a_0r_0}{\sqrt{r^2-r^2_0}}+\\
\frac{b_0}{2}\Big[\sqrt{r^2-r^2_0}\Big\{2c_0(-1+c_0M)+\frac{c^2_0r}{2}+\frac{2M}{r^2}+\frac{2}{r}-\frac{4c_0M}{r}\Big\}\Big].
\end{split}
\label{eq:a24}
\end{equation}
Hence, if there is no mass distribution between Earth and Venus then the total time taken by the pulse to go from Earth to Venus and then comes back to the Earth in $r\gg r_0$ limit is 
\begin{equation}
T_1=2t_1=2\Big[\sqrt{r^2_e-r^2_0}+\sqrt{r^2_v-r^2_0}-a_0r_e-a_0r_v+\frac{b_0}{48}(48+36c^2_0r^2_0)\{Ei(-c_0r_e)+Ei(-c_0r_v)\}\Big],
\label{eq:a25}
\end{equation}
and the time taken by the signal to go from Earth to Venus and returns to Earth in presence of the mass distribution in $r\gg r_0$ limit is
\begin{equation}
\begin{split}
T_2=2t_2=2\Big[\sqrt{r^2_e-r^2_0}+\sqrt{r^2_v-r^2_0}+2M\ln\Big(\frac{2r_e}{r_0}\Big)+2M\ln\Big(\frac{2r_v}{r_0}\Big)+2M+b_0c_0r_e(-1+c_0M)+\\
b_0c_0r_v(-1+c_0M)+b_0-2c_0Mb_0+\frac{b_0c^2_0}{4}(r^2_e+r^2_v)\Big].
\end{split}
\label{eq:a26}
\end{equation}
Hence the excess time due to GR correction and the axion mediated fifth force is 
\begin{equation}
\begin{split}
\Delta T=T_2-T_1=4M\Big[\ln\Big(\frac{4r_er_v}{r^2_0}\Big)+1\Big]+2b_0c_0(-1+c_0M)(r_e+r_v)+\frac{b_0c^2_0}{2}(r^2_e+r^2_v)+2b_0-\\
4c_0Mb_0+2a_0(r_e+r_v)+\frac{b_0}{24}(48+36c^2_0r^2_0)[Ei(-c_0r_e)+Ei(-c_0r_v)].
\end{split}
\label{eq:a27}
\end{equation}
In absence of axion mediated fifth force, $a_0=0,b_0=0,c_0=0$ and from Eq.(\ref{eq:a27}) we get back the standard GR result
\begin{equation}
\Delta T=\frac{4GM}{c^3}\Big[\ln\Big(\frac{4r_er_v}{r^2_0}\Big)+1\Big]=2\times 10^{-4}\rm{sec},
\end{equation}
where we reinsert $G$ and $c$.
\section{Constraints on axion parameters from light bending and Shapiro time delay measurements}
\label{sec5}
The contribution of axions in the light bending must be within the excess of the GR prediction which implies $(\Delta\phi)_{obs}-\Delta\phi_{GR}\geq\Delta\phi_{axions}$. Hence, from Eq.(\ref{eq:a14}) we can write
\begin{equation}
\Delta\phi_{axions}=\frac{\frac{4M}{b^2}+\frac{q_1q_2}{2\pi M_pL^2}(1-0.347m_a^2b^2)}{\frac{1}{b}+\frac{q_1q_2m_a^2b^2}{8\pi M_pL^2}}-\frac{4M}{b},
\label{eq:a28}
\end{equation}
where $q_1=4\pi f_aR_\odot$, $q_2=4\pi f_aR_\oplus$, $L^2=MD(1-e^2)$. The parameters $b\sim R_\odot$ and $R_\oplus$ are the solar radius and Earth radius respectively. $D$ is the semi major axis of Earth's orbit and $e$ is the orbital eccentricity. Now the uncertainty in the measurement of light bending from the GR prediction is $10^{-4}$ which puts upper bound on the axion decay constant $f_a$ from Eq.(\ref{eq:a28}) as 
\begin{equation}
f_a\lesssim 1.58\times 10^{10}\rm{GeV}.
\label{eq:a29}
\end{equation}
Similarly, the contribution of axions in the Shapiro time delay must be within the excess of GR result which yields $(\Delta T)_{axions}$ from Eq.(\ref{eq:a27}) as 
\begin{equation}
\begin{split}
\Delta T_{axions}=2b_0c_0(-1+c_0M)(r_e+r_v)+\frac{b_0c^2_0}{2}(r^2_e+r^2_v)+2b_0-4c_0Mb_0+\\
2a_0(r_e+r_v)+\frac{b_0}{24}(48+36c^2_0r^2_0)
[Ei(-c_0r_e)+Ei(-c_0r_v)].
\end{split}
\label{eq:a30}
\end{equation} 
Now the uncertainty in the measurement of Shapiro time delay from the GR result is $2\times 10^{-5}$ which puts upper bound on the axion decay constant by using Eq.(\ref{eq:a30}) as 
\begin{equation}
f_a\lesssim 9.85\times 10^6\rm{GeV}.
\label{eq:a31}
\end{equation}
Hence, the stronger bound on axion decay constant $f_a$ is obtained from Shapiro time delay. The mass of the axion is constrained by the distance between the Earth and Sun which gives $\frac{1}{D}=m_a\lesssim1.33\times 10^{-18}\rm{eV}$.
\begin{figure}
\centering
\includegraphics[width=4.0in,angle=360]{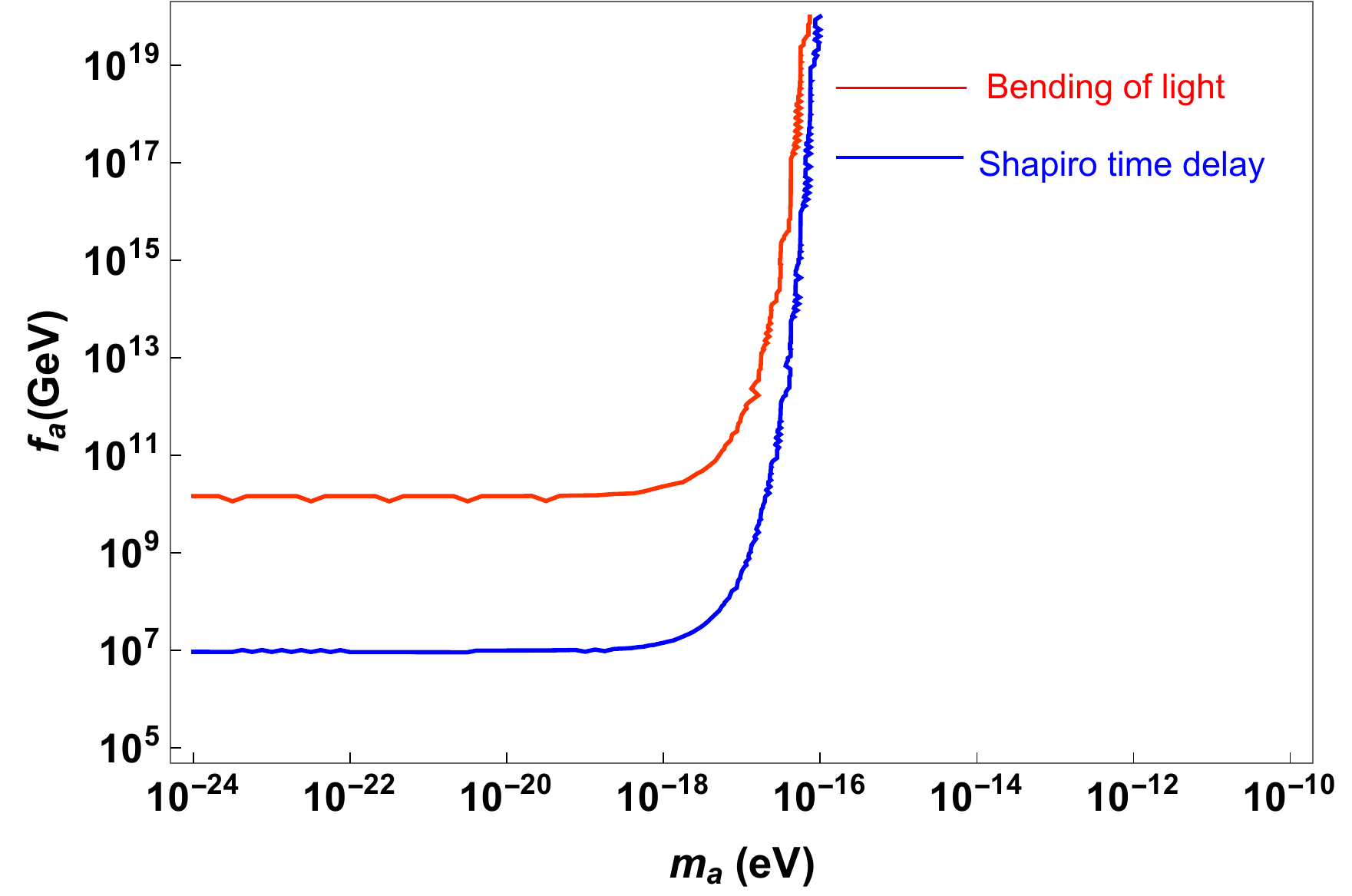}
\caption{Variation of $f_a$ with $m_a$ from light bending and Shapiro delay measurements.}
\label{fig:bending}
\end{figure}
In Fig.\ref{fig:bending} we numerically solve Eq.(\ref{eq:a7}) and Eq.(\ref{eq:a22}) and show the bounds on axion parameters obtained from light bending and Shapiro time delay. The red curve denotes the variation of $f_a$ with $m_a$ for light bending measurement and the blue curve denotes $f_a$ vs. $m_a$ for Shapiro time delay measurement. The region above those curves are excluded. 

We put the upper bounds on the ratio of axionic fifth force to the gravitational force as $\alpha=\frac{q_1q_2}{4\pi Gm_1m_2}\lesssim 10^{-2}$ from light bending and $\alpha=\frac{q_1q_2}{4\pi Gm_1m_2}\lesssim 4.12\times 10^{-9}$ from Shapiro time delay. The Shapiro time delay puts stronger bound on $\alpha$. Hence the axionic fifth force is weaker than the gravitational force by a factor of roughly $10^9$. In Table \ref{tableI} we summarize the bounds on $f_a$ and $m_a$ from light bending and Shapiro time delay.

\begin{table}
\caption{\label{tableI} Summary of axion decay constant ($f_a$) and the ratio of axionic fifth force to gravity ($\alpha$) obtained from light bending and Shapiro time delay for ALPs of mass $m_a\lesssim 1.33\times 10^{-18}\rm{eV}$.}
\begin{tabular}{ lcc  }
 \hline
Experiments & axion decay constant ($f_a$) & $\alpha$  \\
 \hline
Light bending  &  $ \lesssim 1.58\times 10^{10}\rm{GeV}$ & $\lesssim10^{-2}$\\
 Shapiro time delay & $ \lesssim 9.85\times 10^{6}\rm{GeV}$ & $\lesssim4.12\times 10^{-9}$\\
 
 \hline
\end{tabular} 
\end{table}

\begin{figure}
\centering
\includegraphics[width=4.0in,angle=360]{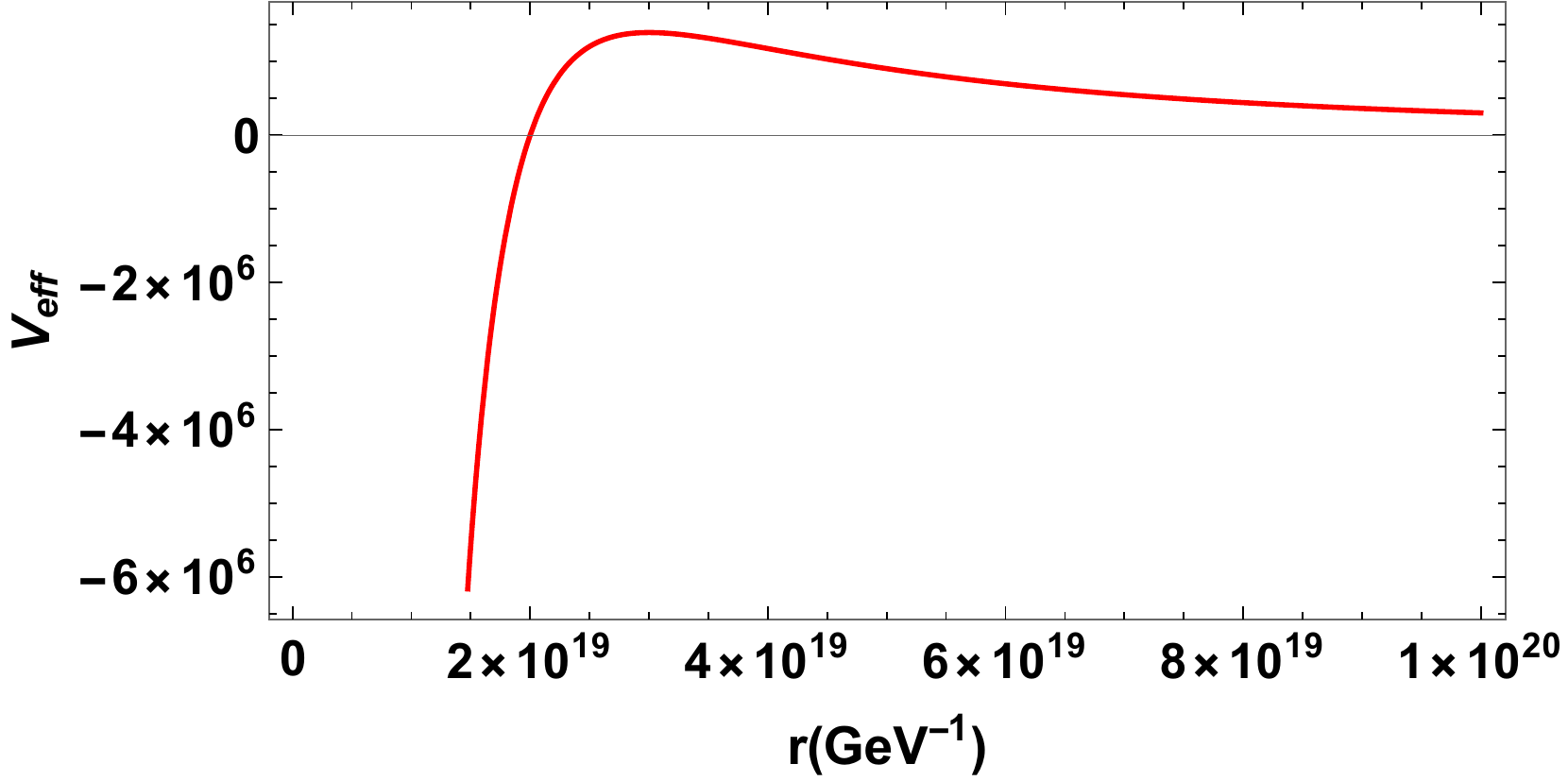}
\caption{Variation of effective potential with distance.}
\label{fig:potential}
\end{figure}

In Fig.\ref{fig:potential} we plot Eq.(\ref{eq:a5}) and show the variation of effective potential with distance. The nature of the potential does not change from its standard GR result in presence of long range axionic Yukawa potential. At $r=3M$, we have circular unstable orbit.
\section{Constraints on axionic fuzzy dark matter from the measurements of light bending and Shapiro time delay}
\label{sec6}
In Sec.\ref{sec2a}, we have discussed that the celestial objects like Sun and Earth can be the sources of ultralight axions or ALPs and they can be possible candidates of FDM whose mass is $\mathcal{O}(10^{-21}\rm{eV}-10^{-22}\rm{eV})$ and has a de Broglie wavelength of order kpc scale. In the begining of the universe, we can write the action of the dynamical axion field as
\begin{equation}
S=\int d^4x \sqrt{-g}\mathcal{L}=\int d^4x \sqrt{-g}\Big[\frac{1}{2}\partial_\mu a\partial ^\mu a-V\Big(\frac{a}{f_a}\Big)\Big],
\label{eq:a32}
\end{equation}
where $g=\det(g_{\mu\nu})$ is the determinant of the metric and the axion field evolves with a periodic potential 
\begin{equation}
V\Big(\frac{a}{f_a}\Big)=m^2_af^2_a\Big[1-\cos\Big(\frac{a}{f_a}\Big)\Big].
\label{eq:a33}
\end{equation}
Using Eq.(\ref{eq:a33}), we can solve the action Eq.(\ref{eq:a32}) to obtain the equation of motion of the axion field in Friedman-Robertson-Walker (FRW) spacetime in Fourier space as
\begin{equation}
\ddot{a_k}+3H\dot{a_k}+\frac{k^2}{R^2}a_k+m_a^2a_k=0,
\label{eq:a34}
\end{equation} 
where $H$ is the Hubble parameter, $R(t)$ is the scale factor in FRW spacetime. In Fourier space, all the modes decouple and for non relativistic or zero mode, we can omit the third term in Eq.(\ref{eq:a34}). Hence, the axionic field has a damped harmonic oscillatory solution. If $H\gtrsim m_a$, then the axion field takes a constant value $a_0=\theta_0f_a$ which fixes the initial misalignment angle $\theta_0$. After that the axion starts oscillating with a frequency $\sim m_a$. The oscilation starts at $H\sim m_a$ and the energy density of the axion field is damped as $\frac{1}{R^3}$. Hence, at late time the axion field varies as $a\propto T^{\frac{3}{2}}\cos({m_at})$, where $T=\frac{1}{R}$ is the temperature of the universe at that epoch and the axion field energy density redshifts like a cold dark matter. With the expansion of the universe, the ratio of the energy densities of dark matter and radiation increases as $\frac{1}{T}$ and at $T\sim 1\rm{eV}$, the dark matter starts dominating over radiation. Hence, the axionic FDM relic density becomes
\begin{equation}
\Omega_{FDM}h^2\sim 0.12\Big(\frac{a_0}{10^{17}\rm{GeV}}\Big)^2\Big(\frac{m_a}{10^{-22}\rm{eV}}\Big)^\frac{1}{2}.
\label{eq:a35}
\end{equation} 
The initial misalignment angle can take values from $-\pi$ to $+\pi$. The coupling of ALPs with matter is proportional to $\frac{1}{f_a}$. Hence, large values of $f_a$ correspond to weaker coupling of axions with matter. The ALPs of mass $m_a\in (10^{-21}\rm{eV}-10^{-22}\rm{eV})$ sourced by Earth and Sun can be the candidate of FDM if $f_a$ is $\mathcal{O}(10^{17}\rm{GeV})$ and $\theta_0\sim\mathcal{O}(1)$. Any value of $f_a$ other than $10^{17}\rm{GeV}$ requires fine tuning of $\theta_0$ which can take values $-\pi<\theta_0<\pi$. From Sec.\ref{sec5}, we obtain the stronger bound on $f_a$ from Shapiro time delay as $f_a\lesssim 9.85\times 10^{6}\rm{GeV}$ and Eq.(\ref{eq:a35}) implies that if the ultralight ALPs have to satisfy FDM relic density, then the ALPs do not couple with nucleons.
\section{Conclusions}
\label{sec7}
In this paper, we have obtained the upper bounds on the axion decay constant from light bending and Shapiro time delay measurements if ALPs contribute to the uncertainty in the measurements of those two experiments. The Shapiro time delay gives the stronger bound on the axion decay constant as $f_a\lesssim 9.85\times 10^6\rm{GeV}$. The sign change of the axion potential due to high nucleon density causes the Sun and the Earth as the possible sources of ALPs. The mass of axion is constrained by the distance between Earth and Sun which gives the upper bound on the mass of axion as $m_a\lesssim 1.33\times 10^{-18}\rm{eV}$. The ultralight nature of axions results a long range Yukawa behaviour of axion field over the distance between Earth and Sun. The presence of long range Yukawa type axion mediated fifth force changes the effective gravitational potential between Earth and Sun and contributes to the time dilation along with the GR effect. The long range axionic fifth force is $10^9$ times smaller than the gravitational force. The upper bounds on $m_a$ and $f_a$ disfavours ALPs as FDM candidates. The paper \cite{Poddar:2019zoe} also disfavors ALPs as FDM from the orbital period loss of compact binary systems. However, the bound on $f_a$ obtained in this work is much stronger than \cite{Poddar:2019zoe}. For single field slow roll inflation, the Hubble scale is $H_*=8\times10^{13}\sqrt{\frac{r}{0.1}}\rm{GeV}$ \cite{Enqvist:2017kzh}, where $r$ is primordial tensor to scalar ratio. The upper bound on $f_a$ that we have obtained in this paper satisfies $2\pi f_a<H_*$ which implies ALP symmetry breaking occurs after inflation. Hence, there will be no constraints on ALPs from isocurvature perturbations. However, the FDM model is in strong tension from Lyman-$\alpha$ forest \cite{Kobayashi:2017jcf,Irsic:2017yje}. The ultralight ALPs in our paper can be probed in the precession measurements of light bending and Shapiro time delay.
\section*{Ackowledgements}
The author would like to thank Professor Subhendra Mohanty for his valuable suggestions and discussions. The author is also grateful to Dr. Soumya Jana for going through the manuscript and providing useful comments.
\bibliographystyle{utphys}
\bibliography{bir}
\end{document}